\documentclass[12pt,preprint]{aastex}

 \usepackage{epstopdf}
 \slugcomment{Accepted Version 1}

\newcommand\pv{\mbox{$p_{V}$}}

\begin{document}

 \DeclareGraphicsExtensions{.pdf,.gif,.jpg}

 \title{Characterizing Subpopulations within the Near Earth Objects with NEOWISE: Preliminary Results}
\author{A. Mainzer\altaffilmark{1}, T. Grav\altaffilmark{2}, , J. Masiero\altaffilmark{1}, J. Bauer\altaffilmark{1,3}, R. S. McMillan\altaffilmark{4}, J. Giorgini\altaffilmark{1}, T. Spahr\altaffilmark{5}, R. M. Cutri\altaffilmark{3}, D. J. Tholen\altaffilmark{6}, R. Jedicke\altaffilmark{6}, R. Walker\altaffilmark{7}, E. Wright\altaffilmark{8}, C. R. Nugent\altaffilmark{9} }

 \altaffiltext{1}{Jet Propulsion Laboratory, California Institute of Technology, Pasadena, CA 91109 USA}
 \altaffiltext{2}{Planetary Science Institute, Tucson, AZ USA}
\altaffiltext{3}{Infrared Processing and Analysis Center, California Institute of Technology, Pasadena, CA 91125, USA}
\altaffiltext{4}{Lunar and Planetary Laboratory, University of Arizona, 1629 East University Blvd., Tucson, AZ 85721-0092, USA}
\altaffiltext{5}{Minor Planet Center, Harvard-Smithsonian Center for Astrophysics, 60 Garden Street, Cambridge, MA 02138 USA}
\altaffiltext{6}{Institute for Astronomy, University of Hawaii, 2680 Woodlawn Drive, Honolulu, HI USA}
\altaffiltext{7}{Monterey Institute for Research in Astronomy, Monterey, CA USA}
\altaffiltext{8}{Department of Physics and Astronomy, UCLA, PO Box 91547, Los Angeles, CA 90095-1547 USA}
\altaffiltext{9}{Department of Earth and Space Sciences, UCLA, 595 Charles Young Drive East, Box 951567, Los Angeles, CA 90095-1567 USA}

 \email{amainzer@jpl.nasa.gov}

 \begin{abstract}
We present the preliminary results of an analysis of the sub-populations within the near-Earth asteroids, including the Atens, Apollos, Amors, and those that are considered potentially hazardous using data from the Wide-field Infrared Survey Explorer (WISE).  In order to extrapolate the sample of objects detected by WISE to the greater population, we determined the survey biases for asteroids detected by the project's automated moving object processing system (known as NEOWISE) as a function of diameter, visible albedo, and orbital elements. Using this technique, we are able to place constraints on the number of potentially hazardous asteroids (PHAs) larger than 100 m and find that there are $\sim4700\pm1450$ such objects.  As expected, the Atens, Apollos, and Amors are revealed by WISE to have somewhat different albedo distributions, with the Atens being brighter than the Amors.  The cumulative size distributions of the various near-Earth object (NEO) subgroups vary slightly between 100 m and 1 km.  A comparison of the observed orbital elements of the various sub-populations of the NEOs with the current best model is shown.     

 \end{abstract}

 \section{Introduction}
Over its 4.5 billion year history, our solar system's planets (including Earth) have experienced many impacts from small bodies such as asteroids and comets.  Evidence exists in the form of craters, fossil records of extraterrestrial materials \citep{Shoemaker.1983a}, and visual evidence of present-day impacts such as Comet Shoemaker-Levy crashing into Jupiter \citep{Levy.1995a} and 2008 TC3/Almahatta Sita disintegrating over Sudan \citep{Jenniskens.2009a}.  NASA's lunar impact monitoring program has detected dozens of impact-generated flashes on the Moon over the past several years \citep{Suggs.2008a}.  

The near-Earth object (NEO) population consists of asteroids and comets with perihelion distances $q\leq1.3$ AU.  Only some NEOs are classified as ``potentially hazardous asteroids" (PHAs); PHAs are formally defined as the subset of near-Earth asteroids (NEAs) whose minimum orbit intersection distance (MOID) with the Earth is 0.05 AU or less and whose absolute magnitude ($H$) is 22.0 or brighter (\emph{http://neo.jpl.nasa.gov/groups.html}).  The potential to make close Earth approaches does not mean that a PHA will impact the Earth.  Although their orbits may intersect, both objects may not be at that point of intersection simultaneously.  Nonetheless, PHAs are automatically assessed for impact probability and their orbit solutions are refined as new measurements become available. For example, the NASA Near Earth Object Program's SENTRY system\footnote{\emph{http://neo.jpl.nasa.gov/risk/}} and the European Union's NEODyS system\footnote{\emph{http://newton.dm.unipi.it/neodys/}} both automatically scan the most current asteroid catalogs to compute impact probabilities over the next 100 years. 

In 2005, Congress assigned NASA the task of detecting 90\% of all NEOs with a diameter larger than 140 m by the year 2020.  The size limit of 140 m in diameter was chosen by \citet{Stokes.2003a} because discovery of 90\% of such NEOs would retire 90\% of the remaining estimated risk integrated over all size ranges.  It represents a trade-off between the increasing numbers of objects with decreasing size, and the decreasing impact energy with decreasing size.  However, objects as small as 30-50 m can inflict damage similar to that of the 1908 Tunguska event \citep{NRC.2010a} or the event that formed the 1.2 km wide Barringer Crater \citep{Melosh.2005a}.  Since previous surveys have predominantly discovered NEOs and PHAs using visible light observations, the size requirement is specified in terms of $H$, assuming that \pv=14\% and the effective spherical diameter $D$ is given by the relationship \citep{Fowler.1992a, Bowell.1989a} \begin{equation}D = \left[\frac{1329\cdot 10^{-0.2H}}{p_{v}^{1/2}}\right].\end{equation} 

Although nearly $\sim$8,800 NEOs have been discovered to date at all size ranges\footnote{\emph{http://neo.jpl.nasa.gov/stats}}, this number represents only a fraction of the total population thought to exist \citep{Bottke.2002a,Mainzer.2011b}.  Of this number, only $\sim$1300 are formally classified as PHAs with $H\leq22$ mag, although if the definition of PHA is expanded to include those objects with $H<25.5$ mag (equivalent to 28 m if \pv\ is assumed to be 14\%), but with a PHA-like orbit (having Earth MOID $\leq$0.05 AU), then the current catalog would include $\sim$2600 PHAs.   However, since most of the surveys that have produced the catalog of currently known NEOs observe at visible wavelengths and with particular observational patterns (spending most of their time looking at opposition), these may not accurately represent the actual fraction of PHAs at all size scales.  Since impact energy scales as the cube of diameter and linearly with density, it is clearly important to understand not just the numbers of NEOs with orbits close to the Earth's but their sizes and compositions as well.  Since some NEOs make many more close approaches to Earth than others, they will also tend to be easier to reach energetically either with robotic or human-crewed spacecraft, furthering their importance as a population of interest \citep{Abell.2009a}.

The potential source regions of the NEOs have been modeled most recently by \citet{Bottke.2002a}.  This work built upon that of previous authors who were able to identify and model successively more source regions in and around the Main Belt as computer processor power increased \citep{Wetherill.1979a, Wetherill.1985a, Wetherill.1987a, Wetherill.1988a, Rabinowitz.1997a, Rabinowitz.1997b, Farinella.1994a, Gladman.1997a, Gladman.2000a}.  The contribution from comets in the \citet{Bottke.2002a} model built upon work by \citet{Levison.1994a, Duncan.1997a} and \citet{Weissman.1996a}.  \citet{Bottke.2002a} made the assumption that the absolute magnitude distribution of the NEOs was independent of orbital elements and source regions and did not make any specific predictions about the subset of the NEOs that are potentially hazardous.  The model also assumed that the forces acting upon NEOs trapped in various source region resonances were independent of asteroid size or albedo.  Given the size of sample of NEOs and PHAs that were known at the time ($\sim$2000 NEOs and several hundred PHAs, according to \emph{http://neo.jpl.nasa.gov/stats/}), these assumptions were reasonable; however, it was noted in that paper that obtaining more accurate number and size distributions of PHAs would be desirable.

Using thermal infrared observations from the \emph{Wide-field Infrared Survey Explorer's} NEOWISE project \citep{Wright.2010a, Mainzer.2011a}, we made a preliminary assessment of the numbers and physical properties of the NEAs larger than 100 m in diameter as a whole \citep[][hereafter M11b]{Mainzer.2011b}. By virtue of being a space-based thermal infrared survey telescope capable of independently discovering new NEOs, NEOWISE provided a sample of NEOs and PHAs that is essentially unbiased with respect to visible albedo (\pv) and inclination (M11b).  In that work, we found that the orbital element model of \citet{Bottke.2002a} yielded a reasonable representation of the population (after survey biases were removed), although some variances were observed.  These are explored in greater detail in this paper.  In M11b, we also determined that there were fewer NEAs larger than 100 m than previous predictions suggested.  Now, we turn our attention to characterizing sub-populations within the NEOs, including the PHAs, in an effort to better understand their origins, evolution, and potential impact hazard. Similar techniques to those described in M11b are used to obtain debiased numbers, sizes, and albedos for these sub-populations.    

\subsection{Definitions and Terminology}
NEOs have traditionally been divided into groups depending on their orbital characteristics, although the choice of definitions for these categories may seem somewhat arbitrary as objects can sometimes evolve from one type to another.  Atens are defined as NEOs with semimajor axis $a <$1.0 AU and aphelion distance $Q \geq$0.983 AU.  Apollo-type NEOs have $a \geq$1.0 AU and $q \leq$1.017 AU. Thus, both have orbits that cross the Earth's, but an Aten orbit lies mostly inside of the Earth's orbit, and an Apollo orbit lies mostly outside.  Amors have 1.017 $< q \leq$1.3 AU; these objects are currently on orbits that do not cross the Earth's.  As will be discussed in greater detail below, some Amors are likely to have very different origins than the rest of the NEOs.  Objects with orbits that lie entirely within Earth's are known either as interior-to-Earth objects (IEOs) or Apoheles.  All NEOs are thought to be transient, with dynamic lifetimes of  $10^6$ to $10^8$ years \citep{Morbidelli.1998a}, depending on their place of origin.  Asteroids and comets are known to transition between the above-described categories.  For example, (99942) Apophis will change from an Aten to an Apollo after its 2029 close approach to the Earth \citep{Tholen.2012a}.  Dynamically, Atens and Apollos are closely related.  

The Amors may be subdivided into two groups, those with perihelia very close to Earth's, and those with larger perihelia. A fundamental dividing line in the Amor group occurs at around 1.1 AU perihelion.  Objects in Jupiter-crossing orbits can be perturbed inward to as close as $\sim$1.1 AU; however, if the Jupiter encounter velocity is high enough to scatter the object inside of 1.1 AU, then a similar encounter scattering in the opposite direction would result in the object escaping the solar system.  Hence, Jupiter can scatter active or dormant comets inward to $\sim$1.1 AU, but in general not further without additional perturbations.  It would therefore be reasonable to divide the Amors into those with perihelia above and below 1.1 AU when studying their physical and orbital characteristics.    

As discussed above, PHAs have a minimum Earth orbit-to-orbit separation (Earth ``MOID")  $\leq$0.05 AU \citep{Sitarski.1968a,Gronchi.2005a}.  The magnitude limit is a proxy for inferring an object of a diameter greater than $\sim$140 meters, while the MOID gives the closest Earth-approach possible based on the two instantaneous osculating ellipse geometries. An actual encounter prediction considers the location of both objects; the MOID does not, considering only their osculating orbital elements.  The MOID is not constant, but can vary secularly over time due to orbit  precession and other perturbation dynamics. This rate of change is generally less than 0.05 AU per century \citep{Bowell.1996a}, while asteroid orbit solutions, based on typical ground-based optical astrometry from a single discovery apparition, can be validly extrapolated on average about 80 years into the future \citep{Ostro.2004a}. An object currently having a MOID $>$0.05 AU is therefore unlikely to dynamically evolve into a potential impactor within the forseeable future.  The PHA definition therefore serves to filter those objects large enough to penetrate the Earth's atmosphere, while also having an orbit that could in principle dynamically evolve to an Earth-crossing MOID near zero within the predictable future of at least the next 100 years.

\section{Discussion}

\begin{figure}
\figurenum{1}
\includegraphics[width=6in]{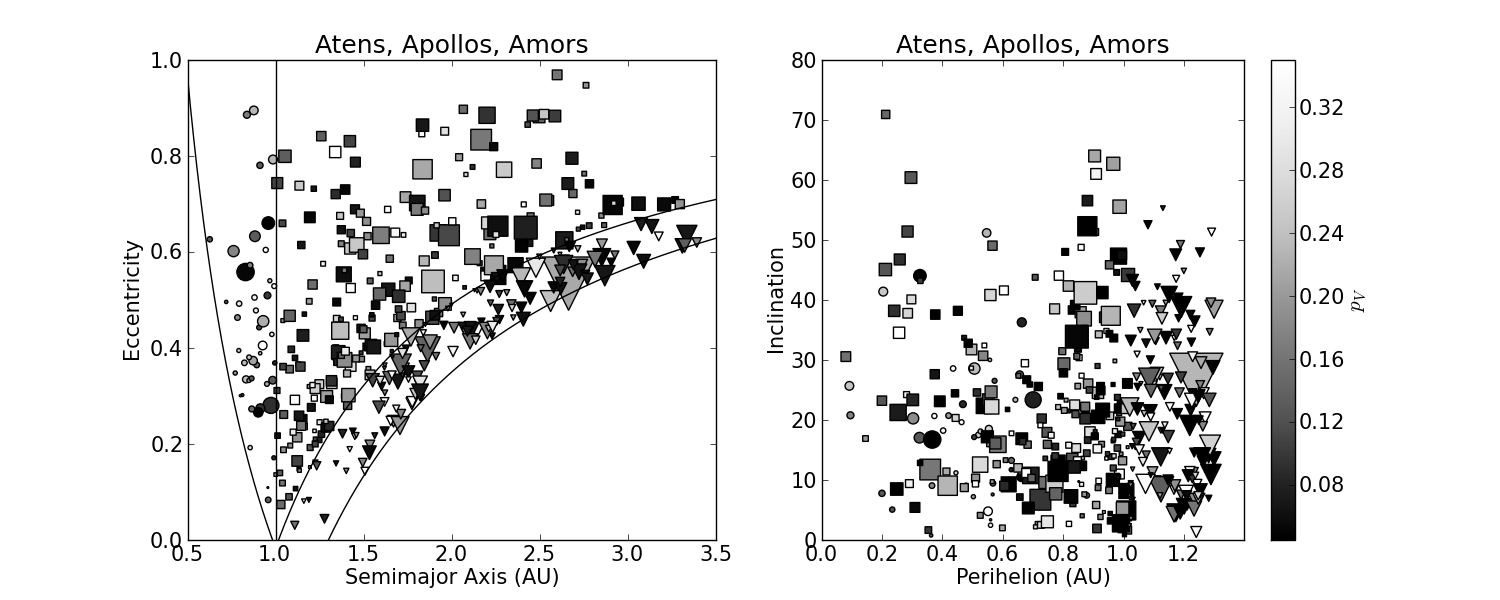}
\caption{\label{fig:orbital_elements_allNEAs}Orbital elements of the 429 NEOs observed by NEOWISE during the fully cryogenic portion of the mission.  Symbol sizes are proportional to the objects' spherical equivalent diameters, and the grayscale is proportional to \pv.  Amors are shown as triangles; Apollos are shown as squares, and Atens are circles. These observed distributions have not had survey biases removed; however, as shown in M11b, the NEOWISE survey is largely unbiased with respect to \pv.  The preponderance of low albedo objects among the Amors and higher albedo objects among the Atens is therefore likely representative of the true populations. However, the fact that the Amors appear systematically larger may be due to observational bias: Amors tend to spend most of their time at larger heliocentric distances compared to the Atens, so they must be larger to have been detected by NEOWISE. It is necessary to debias the survey in order to determine the true size distribution of the populations.}
\end{figure}

\begin{figure}
\figurenum{2}
\includegraphics[width=6in]{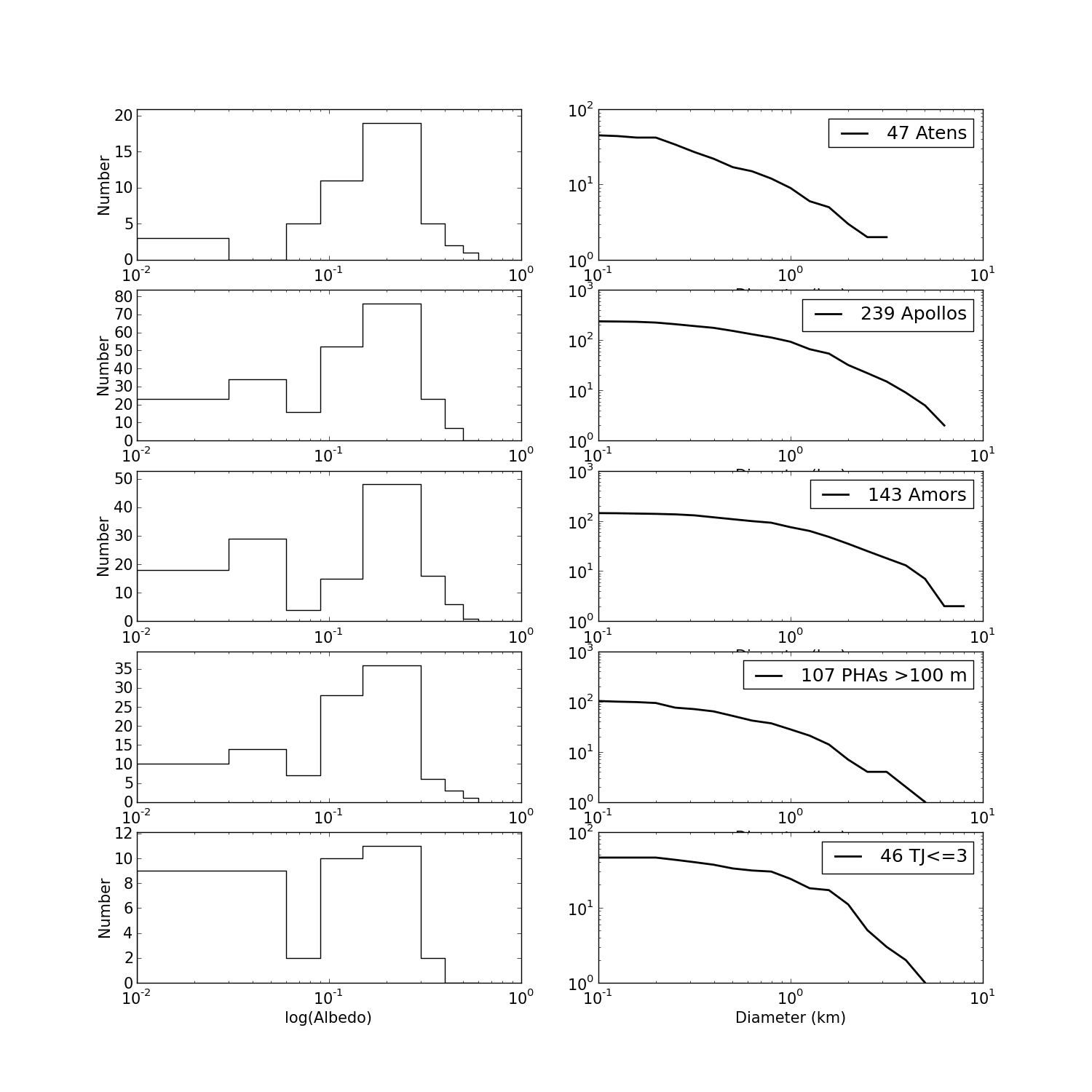}
\caption{\label{fig:neo_types1}Visible albedo and cumulative size distributions for the Atens, Apollos, Amors, low MOID, and low Jupiter Tisserand objects observed during the fully cryogenic portion of the NEOWISE mission. These observed distributions are used as the basis for modeling the underlying populations.}
\end{figure}

Figure \ref{fig:orbital_elements_allNEAs} shows the orbital element plots for the 429 NEAs detected by NEOWISE during the fully cryogenic portion of the mission, along with a graphic representation of the effective spherical diameter and visible albedo of each object as determined from the application of the Near Earth Asteroid Thermal Model \citep[NEATM;][]{Harris.1998a} to the WISE data. M11b gives a description of the thermal modeling for the NEAs.  The fully cryogenic portion of the mission was the phase during which sufficient solid hydrogen remained to maintain all four WISE bands at their operational temperatures. After a seven-week period during which bands W1, W2, and W3 remained operational, the remaining cryogen was depleted, and the survey continued with only bands W1 and W2.  Because the survey sensitivity and biases changed substantially after the cryogen was depleted, we have only included data from the fully cryogenic portion of the survey in this preliminary analysis. Detailed descriptions of the data reduction and thermal modeling can be found in M11b; see the WISE Explanatory Supplement \citep{Cutri.2011a} for descriptions of the single-epoch image processing, and M11a for a description of the NEOWISE pipeline that extracted moving objects from the WISE data.  The 429 NEAs shown in Figures \ref{fig:orbital_elements_allNEAs} include 116 new discoveries attributed to WISE at the time of publication.  Figure \ref{fig:neo_types1} shows the observed albedo and size distributions for several subpopulations within the NEOs.  As in M11b, we limit our analysis only to objects that are not known to be active comets, although some could be dormant comets (a more detailed discussion of this is given below).  
 
Some trends are apparent from Figures \ref{fig:orbital_elements_allNEAs} and \ref{fig:neo_types1}.  First, NEOWISE did not detect any IEOs since WISE only observed near $\sim90^{\circ}$ solar elongation and was essentially unable to view objects inside the Earth's orbit. Second, the Amors appear noticeably larger and darker than the the other NEO subgroups.  This result suggests that there is a correlation between albedo and orbital elements.  We note that in Figure \ref{fig:neo_types1}, the survey biases have not yet been removed; in the sections that follow, we discuss how we determine and account for the survey biases.  By sorting the NEAs into several general populations by their orbital elements and treating each population separately, we can refine our estimates of the numbers, sizes, and orbital element distributions for the NEAs that were produced in M11b.

We will consider separate population models for four distinct classes of NEAs: the Atens, Apollos, Amors, and objects with both a MOID $\leq$0.05 AU and effective spherical diameters larger than 100 m. The latter population encompasses the PHAs but includes additional smaller objects, some of which may have $H\leq$22 mag.  Since WISE thermal infrared measurements do not significantly constrain $H$, we adopt a limiting diameter of $\geq$30 m so as to include objects having PHA-like orbits but which are smaller than the 140 m of the formal definition.  The estimated minimum size of the Tunguska impactor is $\sim$30 m, approximately the size of the smallest object in our sample so far.  Such 30-140 m objects having a MOID $\leq$0.05 AU are referred to as Small PHAs (SPHAs), although the results presented here are limited to those $\geq$100 m, since only four objects below this size were observed.  These four objects are not enough to significantly constrain the form of the cumulative size frequency distribution (SFD) below 100 m (c.f. M11b).  At or near this size limit, the observational biases in the NEOWISE survey become very large, and small changes in the number of detections can therefore produce large changes in the predicted total number of objects in the population at large.  

One such source of uncertainty in the number of detected objects are objects detected by NEOWISE with short observational arcs.  As described in M11b, NEOWISE NEO candidates required ground-based follow-up astrometry within approximately two weeks of their detection by the spacecraft as the average orbital arc from WISE alone was $\sim$1.5 days.  As with M11b, in this preliminary study, we do not include uncertainties due to the objects lacking measured albedos, or the uncertainties introduced by objects with observational arcs short enough to cause large uncertainties in their orbital parameters.  We can make estimates of what fraction of the total sample of 429 NEOs detected by WMOPS in the fully cryogenic survey these objects represent, however.  

For objects with WISE observations but no visible light detections, it is possible to compute a diameter from the thermal measurements alone, but not \pv. Of the 47 Atens detected so far in the fully cryogenic sample, only one received a designation but no optical follow-up (2010 DJ$_{77}$; although its observational arc was sufficient to produce only a small error in its heliocentric distance and therefore its effective diameter).  The current fully cryogenic sample also includes 239 Apollos, 143 Amors, and 107 PHAs (including the SPHA 30-140 m category).  There are 46 NEAs with low Jupiter Tisserand parameters ($T_{J}$). Of the 239 Apollos, six objects were discovered by NEOWISE but have not yet received ground-based visible follow-up, so they have computed diameters, but no albedos: 2010 CA$_{55}$, 2010 DG$_{77}$, 2010 EN$_{44}$, 2010 MA$_{113}$, 2010 MY$_{112}$, and 2010 MZ$_{112}$.  Of the 143 Amors, four have not received optical follow-up: 2010 AQ$_{81}$, 2010 AU$_{118}$, 2010 JA$_{43}$, and 2010 MB$_{113}$.  Out of the entire sample of 429 NEOs detected by WMOPS during the fully cryogenic portion of the mission, there are 22 objects that have designations but arc lengths shorter than 10 days; these objects' orbits could change significantly with additional follow-up. There were $\sim$20 NEO candidates that appeared on the Minor Planet Center's NEO Confirmation Page but received neither follow-up nor designations.   Furthermore, some short-arc objects received designations as non-NEOs, but may be redesignated as NEOs after receiving follow-up; one such example is 2012 CC$_{17}$, which was first observed by WISE in 2010 and designated a Main Belt asteroid.  This object subsequently received follow-up in 2012 and was reclassified as an Amor.  These short-arc objects represent a source of uncertainty that is difficult to quantify, but it is likely that they represent a relatively small fraction of the total sample (about 10-15\%) and so are unlikely to change the results of these population studies dramatically.

Figure \ref{fig:neo_types1} shows the observed, ``undebiased" differences in the properties of these populations.  The observed Atens and Apollos have proportionately more high albedos than the Amors and low $T_{J}$ NEOs.  Of the Atens, PHAs, Apollos, Amors, and NEAs with $T_{J}\leq3$ in the present NEOWISE sample, 17\%, 29\%, 30\%, 35\%, and 46\% respectively of the populations have \pv$<$0.09. The various subpopulations in the sample also have different observed cumulative size distributions, and these may represent real differences in the populations.  However, an alternate explanation for the observed size differences could be that more distant objects had to be larger for WISE to detect them.  In order to distinguish between these two possibilities (either the differences in the observed cumulative size distributions are real or are merely the result of observational biases), the survey biases must be computed and accounted for first.  But as shown in M11b and demonstrated again later in this paper, the NEOWISE survey is relatively unbiased with respect to \pv, so the albedo distributions shown in Figure \ref{fig:neo_types1} are actually representative of the various sub-populations, meaning that the Atens and PHAs do in fact consist of more bright objects than the Amors.  We demonstrate that the NEOWISE survey is essentially unbiased with respect to \pv\ in the following sections.  We note that no correlation between \pv\ and diameter was found in M11b for the NEAs.  However, Figure \ref{fig:neo_types1} indicates that there are correlations between \pv, diameter, and orbital elements.

\subsection{Potentially Hazardous Asteroids}
We first consider the PHA population, extended to include smaller objects $\geq$100 m.  The basic technique that we use in this work and in M11b as well as \citet{Grav.2011b, Grav.2011c} is to numerically determine the NEOWISE survey biases and remove them from the observed sample in order to quantify the characteristics of the population as a whole.  We can determine the survey biases in semimajor axis ($a$), eccentricity ($e$), inclination ($i$), \pv, beaming parameter $\eta$, infrared albedo $p_{IR}$ (see M11b for a more detailed explanation of this quantity), and $D$ through creation of synthetic populations.  First, the sensitivity to moving objects as a function of WISE magnitude was found by computing the fractions of previously known objects with well-known orbits that were detected in each frame as a function of galactic and ecliptic coordinates.  Next, the biases in $a$, $e$, $i$, \pv, $\eta$, $p_{IR}$, and $D$ were determined by running the synthetic population through a frame-by-frame model of the NEOWISE survey's fully cryogenic portion. The survey biases are determined by dividing the objects found in each frame of the synthetic survey by the total population of synthetic objects.  These biases are then divided out of the observed distribution of objects detected by NEOWISE following the methods described in \citet{Spahr.1998a} in order to obtain the entire population's characteristics.  The method is a purely numerical Monte Carlo technique, whereas the method of \citet{Jedicke.1998a} models the survey biases analytically.  The procedure we have just described is the general technique for computing the total numbers of the population using only the objects that were either detected or discovered by the NEOWISE moving object detection pipeline.  

Our specific goal for the PHAs was to determine their numbers (down to $\sim$100 m in diameter) and to describe the function of their cumulative size frequency distribution above the 100 m cutoff.  We used the NEOWISE detections and our estimates of the survey's biases to compute the total number of PHAs above 100 m.  As discussed in the previous section, the survey biases at 100 m become very large, and the ability to reliably determine the bias, bias errors, and the errors in the number of detected objects becomes increasingly difficult. We estimated the number of PHAs larger than 1 km that remain undiscovered by dividing the total number of the objects discovered by NEOWISE by our estimate of the survey bias in diameter.  Dividing the total number larger than 1 km that remain to be discovered by the total in the entire population gives an estimate of the current survey completeness.  The precise methodology for determining the survey biases and removing them is described below.  

In order to determine the detection probability as a function of orbital elements and thermophysical properties for asteroids with diameters larger than 100 m, $P(a,e,i,\pv, p_{IR}, \eta, D)$, we constructed a model of the NEOWISE survey performance.  We first determined the completeness with which asteroids with well-known orbits (i.e. numbered objects) were detected by WISE as a function of ecliptic and galactic latitude/longitude as shown in M11b.  As discussed in M11a and M11b, the WISE Moving Object Processing System (WMOPS) required a minimum of five detections in order to link a tracklet.  We created a synthetic population of PHAs using the synthetic solar system model (S3M) from \citet{Grav.2011a}, which uses the orbital element distribution of \citet{Bottke.2002a} as a basis for its NEO population model.  MOIDs were computed for each of the $\sim$268,000 synthetic NEOs in the S3M model using the same methodology used to compute MOIDs for all NEOs by the JPL NEO Program Office. The objects with MOIDs $\leq$0.05 AU were selected. The field centers of each of the millions of pointings in the WISE survey were used to determine whether or not a particular synthetic object would have passed through the WISE field of view for each frame. Finally, NEATM was used to compute fluxes for each synthetic object in each WISE frame; if the flux exceeded the sensitivity threshold and the object met the rules for detection by WMOPS (including number of detections and on-sky velocity; see M11b for details), the synthetic object was counted as ``found" by the simulated survey. The survey biases in $a$, $e$, $i$, \pv, $p_{IR}$, $\eta$, and $D$ were found by dividing the objects found by the simulated survey by the total synthetic population.  

Once the synthetic survey was constructed, we next considered the population detected by the WMOPS pipeline.  NEOWISE detected a total of 107 PHAs during the fully cryogenic portion of the mission. We omitted from consideration any objects with diameters $<$100 m and having arc lengths shorter than three days, as the uncertainty in their orbital elements and MOIDs are so large as to be unable to tell whether or not they really are PHAs; this constraint eliminated one potential PHA larger than 100 m (2010 DG$_{77}$).  The uncertainty caused by whether or not 2010 DG$_{77}$ is a PHA means that we must regard our total population estimate larger than 100 m as a lower limit. However, 2010 DG$_{77}$ represents one object out of a total sample of 102 objects with arc lengths longer than three days and diameters larger than 100 m, so the uncertainty that it produces should be relatively small. We did not consider PHAs smaller than 100 m since NEOWISE detected only four PHAs in this size range, not enough for a statistically meaningful constraint on the population. 

The total number of PHAs larger than 100 m is given by \begin{equation}N_{T}(a,e,i,\pv, p_{IR}, \eta, D>100 m) = \sum_{D=0.1 km}^{10 km} \frac{N_{o}(a, e, i, \pv, p_{IR}, D)}{P_{N}(a, e, i, \pv, p_{IR}, \eta, D)}, \end{equation} where $N_{T}$ is the total number, $N_{o}$ is the number observed by NEOWISE, and $P_{N}$ is the NEOWISE survey bias.  We created a synthetic population of PHAs in order to determine the survey biases in \pv, $D$, $\eta$, and orbital elements for objects with effective spherical diameters as small as 100 m.  We generated 24 sets each containing 5000 synthetic PHAs with physical parameters randomly assigned according to the distributions of \pv, $\eta$, $p_{IR}$, and $D$ observed for the PHAs.  The synthetic population was run through all the frames in the synthetic survey; fluxes were computed for each object, and the survey bias $P_{N}(a, e, i, \pv, p_{IR}, \eta, D)$ was found by comparing the ratio of objects found by the simulated survey to the entire synthetic population.   In other words, $P_{N} = \sum_{D=0.1 km}^{10 km} \frac{S_{found}}{S_{sim}}$.

\begin{figure}
\figurenum{3}
\includegraphics[width=6in]{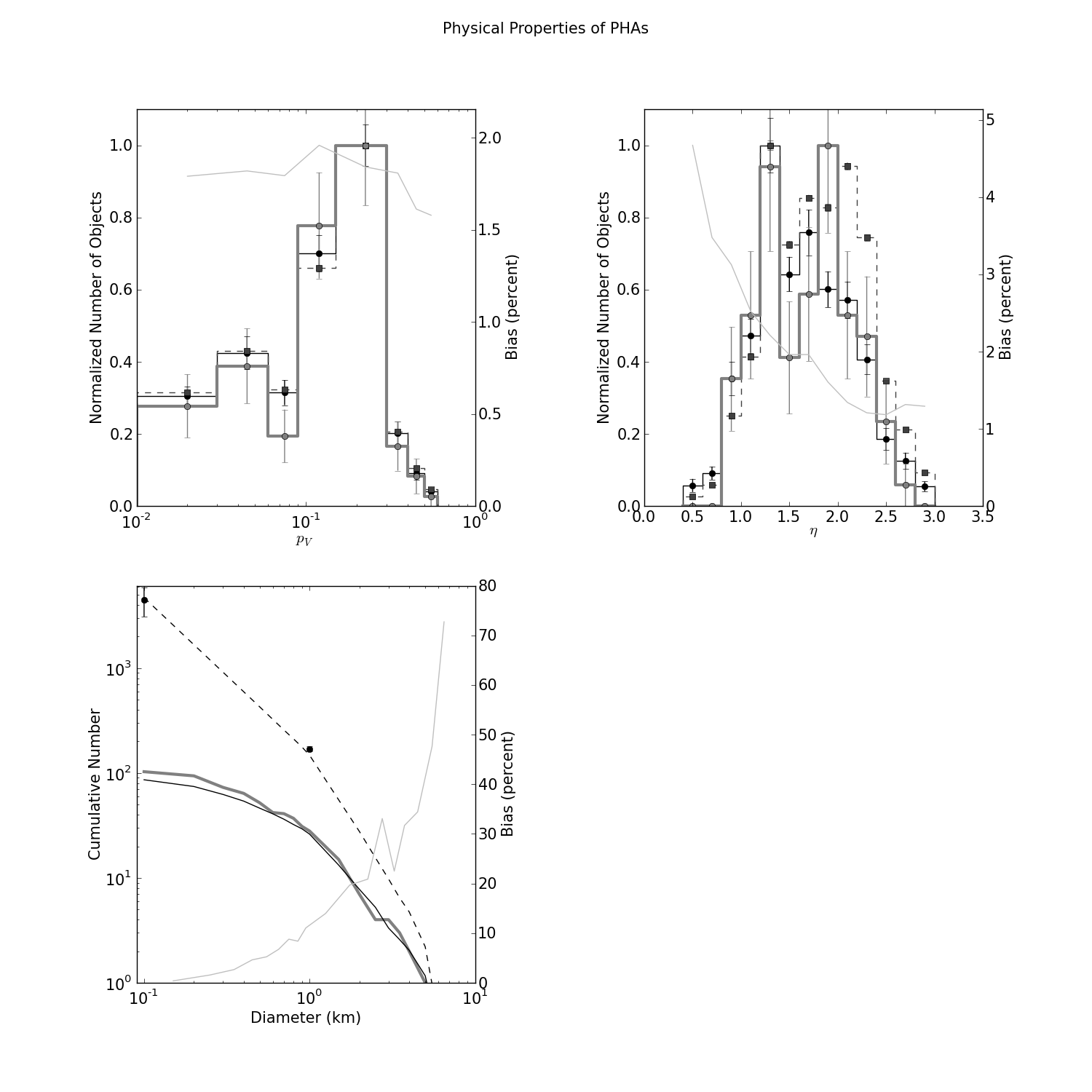}
\caption{\label{fig:debiased_phys_pha} The physical properties (\pv, $\eta$, and $D$) of the entire synthetic population of PHAs larger than 100 m (dashed lines) are compared to the objects that were found by the simulated survey (black lines).  The PHA sample observed by NEOWISE during the fully cryogenic portion of the mission is shown as medium gray lines.  The survey biases (light gray lines) are determined by dividing the distributions of objects found by the simulated survey by the entire synthetic population distributions. The cumulative diameter distribution of the PHAs is represented by a power law with a slope of $\alpha=1.50\pm0.20$ for $D<1$ km.}
\end{figure}

By dividing the observed cumulative size distribution $N_{o}$ by the bias $P_{N}$ and summing over diameter, we computed the total number of PHAs larger than 100 m, $N_{T}$. Errors in the bias were derived from Monte Carlo trials of the synthetic and ``found" populations, and errors in the observed population's cumulative size distribution were assumed to be Poissonian.  Using Equation 3 with the unmodified \citet{Bottke.2002a} orbital elements, we found that there are $N_{T}=\sim4700\pm1450$ PHAs larger than 100 m and 160$\pm$60 PHAs larger than 1 km.  A least-squares minimization was used to fit the PHA cumulative size distribution to a broken power law $N\propto D^{-\alpha}$.   The PHA cumulative size distribution is best fit by a power law with a break at 1.0 km and a slope $\alpha=1.50\pm0.20$ for $D<1.0$ km and $\alpha=2.67\pm0.70$ above 1.0 km.  The slope of the PHAs below 1.0 km reported here (Figure \ref{fig:debiased_phys_pha}) is somewhat steeper than the slope of $\alpha=1.32\pm0.14$ found for the NEAs in M11b, but it is consistent within the error bars.  

We can make an estimate of the number of large ($D>1$ km) PHAs that remain to be discovered by examining the NEOWISE discoveries.  The survey biases of the objects discovered by NEOWISE, $P_{N_{disc}}$, depend not only on the biases of the WISE survey but also on the performances of the other ground-based visible light surveys: these other surveys determine whether or not an object was previously known, i.e. $P_{N_{disc}} = P_{N_{disc}}(P_{N}, P_{vis})$. In order to determine the total survey bias for the new discoveries, it is necessary to find the survey biases for those surveys as well as NEOWISE, $P_{vis}(a, e, i, \pv, D)$.  However, it is likely that most of the large ($D>1$ km) PHAs have already been discovered; out of the 27 PHAs that NEOWISE discovered, only one was larger than 1 km, 2010 LG$_{64}$.  In this case, we can make the approximation that the survey biases of the NEOWISE-discovered PHAs are the same as the survey biases of \emph{all} PHAs larger than 1 km that were detected by NEOWISE: $P_{N_{disc}} \rightarrow P_{N}$.  We can make this approximation because the orbital elements of the discovered population are very similar to those of the entire population in the limit that nearly all objects have been discovered. We use this approximation to estimate the total number of PHAs larger than 1 km that remain to be discovered.  Using Equation 2 to divide the number of PHAs larger than 1 km discovered by NEOWISE (there is only one such object) by the survey bias, we estimate that there are $7\pm7$ large PHAs remaining to be found.  Since only one PHA larger than 1 km was discovered by NEOWISE during the fully cryogenic survey (2010 LG$_{64}$), the resulting number has a large uncertainty. As $P_{N}$ approaches zero, it becomes impossible to place meaningful constraints on the population.  This result ($7\pm7$ PHAs with $D>1$ km remaining to be discovered) is consistent with zero, and the large error bar is a direct consequence of the fact that NEOWISE only discovered one PHA larger than 1 km.  Nevertheless, the implication is that it is likely that few extremely large PHAs remain to be discovered at present.   

We have now determined an estimate of the number of PHAs larger than 1 km ($N_{known}=160\pm$60); the uncertainty derives from both the size of the observed sample as well as our knowledge of the survey biases.  We also have an estimate of the number of large PHAs remaining to be found (7$\pm$7); we can use both of these quantities to estimate the survey completeness.  By dividing the total number found to date by the total population, we approximate that the survey completeness for 1 km and larger PHAs is therefore $\sim$96$^{+3}_{-7}$\%, which is consistent with the 93\% completeness found in M11b for the general NEA population larger than 1 km.  

We cannot use the NEOWISE discoveries to estimate the number of PHAs larger than 100 m that remain to be discovered unless we also compute the survey biases of the visible light surveys, $P_{other}$.  At 100 m, only a small fraction of the total population has been discovered, so the approximation that $P_{N_{disc}} \approx P_{N}$ no longer holds.  In the future, we will attempt to estimate and incorporate the survey biases of the visible light surveys to obtain a more precise answer.  An alternate approximation can be made by turning to the $\sim$8,800 known objects in the literature and applying an estimated albedo distribution in order to convert their $H$ values into $D$ using Equation 1.  To that end, we compiled all the diameters and albedos we could find from the literature from a variety of sources (radar, in situ spacecraft measurements, stellar occultations, and infrared measurements, including those from WISE and the \emph{Spitzer Space Telescope}).  We make the assumption that the albedo distribution drawn from the PHAs with albedo measurements can be applied to the remainder of the objects that do not have albedo measurements.  This assumption may not be entirely appropriate, since the NEOWISE survey is considerably more sensitive to low albedo objects than the visible light surveys that have discovered nearly all of the PHAs, and since objects with low $H$ are much more likely to have higher albedos.  Even the ensemble of albedos derived from observations with \emph{Spitzer} is biased against low albedo objects, since the sample is drawn from the visible light surveys.  Nevertheless, it is safe to regard the albedo distribution that results from the combination of the literature albedos with the WISE albedos as representing the darkest that the distribution can possibly be.  Since we have shown that the NEOWISE survey was more or less unbiased with respect to albedo, the NEOWISE albedo distribution is representative of the actual population.  The known PHAs with low $H$ must therefore be brighter than the NEOWISE sample, and the difference between the known objects' albedos and the NEOWISE albedo distribution will increase as $H$ decreases.  Therefore, the number of PHAs larger than a certain $D$ that one finds by assuming their albedo distribution is represented by an aggregate of WISE and literature albedo measurements should be regarded as  an upper limit.  By taking this net albedo distribution and applying it through many Monte Carlo trials to the known PHAs, we estimate that $\lesssim$1400 PHAs larger than 100 m have been discovered to date out of the $\sim4700\pm$1450 we expect to exist overall.  Hence, we estimate that $\lesssim30\%$ of the PHAs larger than 100 m have been discovered to date.

A similar technique was used in M11b to estimate the number of previously known NEAs larger than 1 km.  However, this estimate should be regarded more as an upper limit, since we now recognize that the albedo distribution given by combining literature measurements plus WISE is likely to be slightly darker than the actual distribution of NEAs with brighter $H$ magnitudes.  $H$ is correlated with albedo; objects with lower $H$ values are more likely to have high albedos.  An improved estimate of the total number and the number remaining to be discovered can be obtained by breaking the NEAs into three separate populations, the Atens, Apollos and Amors, and considering each separately using Equation 2.  As shown in Figure \ref{fig:orbital_elements_allNEAs} and \ref{fig:neo_types1}, there is a correlation between albedo and orbital elements which was not taken into account in the analysis in M11b; in that work, all NEOs were assumed to have the same size and albedo distribution, regardless of their orbital elements.  By separating the NEOs into three sub-populations, we can account for this correlation.  The results of this analysis are discussed below, along with an updated estimate of the numbers of large NEAs.  

\begin{figure}
\figurenum{4}
\includegraphics[width=6in]{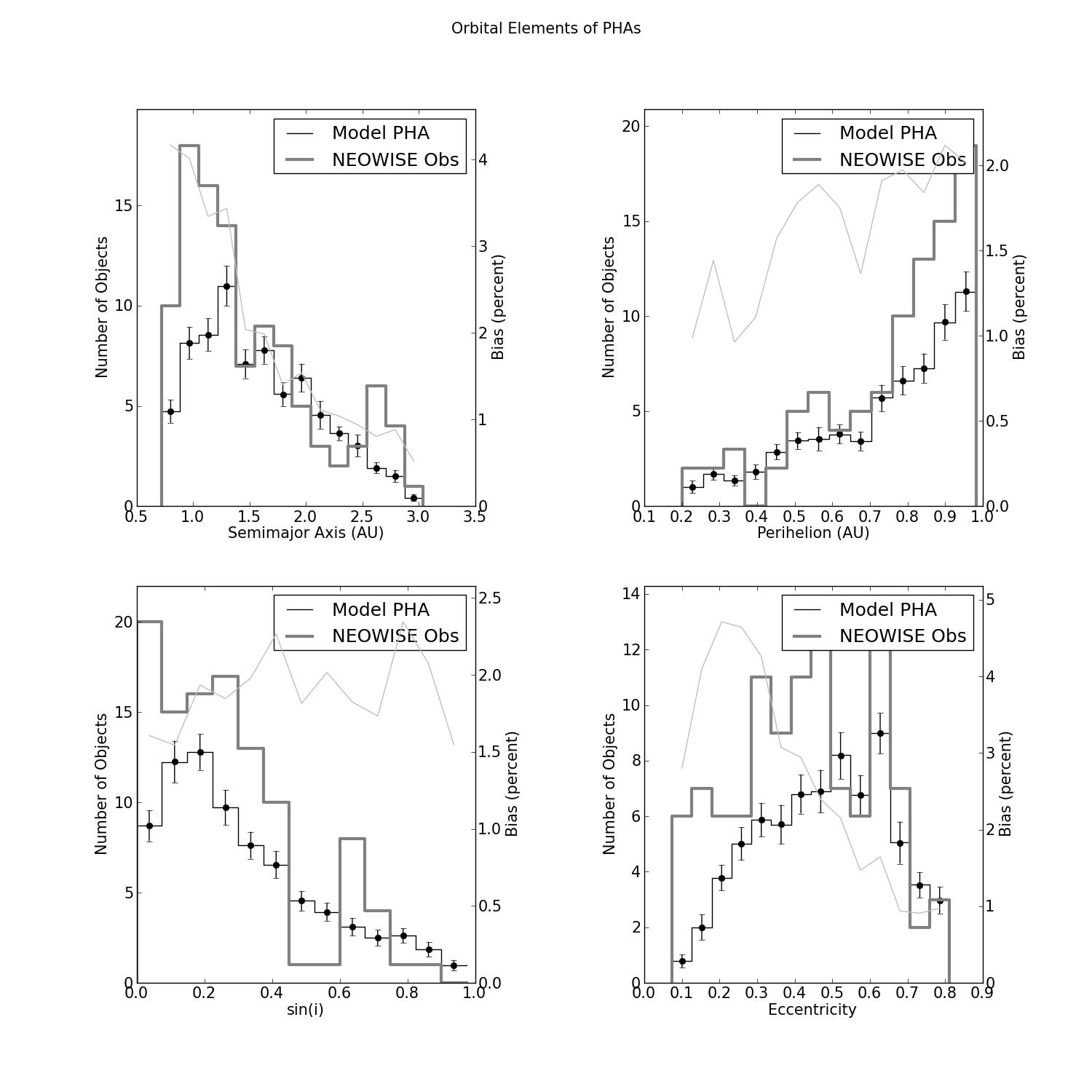}
\caption{\label{fig:debiased_aei_pha} As described in the text, the S3M model under predicts the number of low-inclination PHAs by a factor of 2.3 with a confidence of 99.93\%. The model also over-predicts the number of PHAs between 27-37 degrees of inclination.  Each panel shows the orbital elements of the PHAs observed by NEOWISE during the fully cryogenic portion of the mission (medium gray lines) compared to the number predicted by computing the survey biases (light gray lines) and applying them to the model population predicted by the S3M model (black lines).  Error bars on the model were determined through Monte Carlo simulations.}
\end{figure}

We now consider the orbital elements of the PHAs. Although in M11b we found generally good agreement between the \citet{Bottke.2002a} orbital element model and the overall population of NEOs observed by NEOWISE, some mismatches between the orbital elements of the model population and the observed PHAs are evident (Figure \ref{fig:debiased_aei_pha}). As noted in M11b, the inclination distribution of the NEOWISE-observed NEOs matches that of the S3M model reasonably well.  However, when the inclination distribution of the NEOWISE-observed PHAs is compared with the S3M model, some discrepancies are apparent.  Figure \ref{fig:debiased_aei_pha} shows that the NEOWISE survey is relatively unbiased with respect to inclination, as would be expected from an all-sky survey.  The number of PHAs predicted by the S3M model as a function of the various orbital elements was determined by selecting 10 randomly chosen populations of 20,500$\pm$3000 NEAs (based on our best estimate of the number of the number of NEAs larger than 100 m from M11b) and selecting the PHAs from each trial population.  The biases in the orbital elements were applied to the mean model distribution, leading to a prediction of the number of objects that should have been observed as a function of $a$, $e$, and $i$. There are 20 NEOWISE-detected PHAs with $\sin(i)<0.1$; however, the model predicted that 8.7$\pm$0.9 PHAs should have been detected.  Assuming that the number of PHAs detected by NEOWISE follows a Poisson probability distribution, we conclude that the model underpredicts the number of low-inclination PHAs by a factor of 2.3 with a confidence of 99.94\%, or about 3.5$-\sigma$.  Also, a dearth of observed objects between $\sim27-37^{\circ}$ inclination can be seen in Figure \ref{fig:debiased_aei_pha}.  Two such objects were detected by NEOWISE, but the model predicts 8.4$\pm$0.8; the model therefore overpredicts the number of PHAs with these inclinations by a factor of 4.2 with a confidence of 99.0\%, or about 2.5$-\sigma$.  It is possible that these results indicate that certain source regions may be more important than others.  

\begin{figure}
\figurenum{5}
\includegraphics[width=3in]{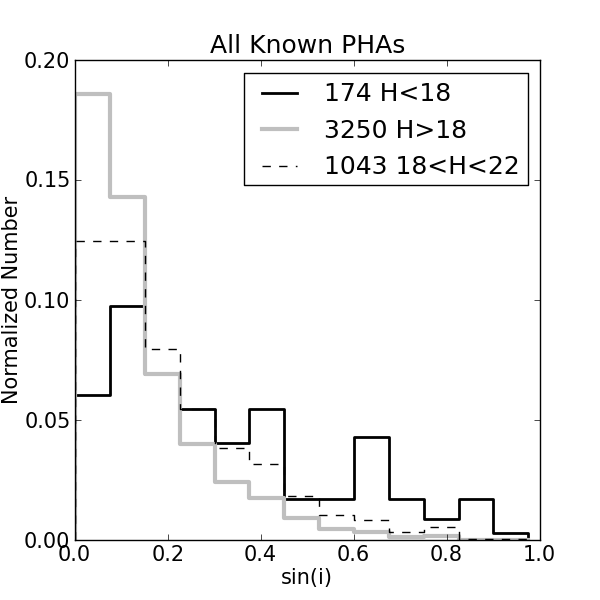}
\caption{\label{fig:known_pha_incl_hist} The known PHAs can be broken down by their $H$ magnitudes when considering their inclination distributions; the sample of PHAs with $H<18$ mag is thought to be roughly complete.  The inclination distribution of these bright PHAs matches the S3M model reasonably well (see Figure \ref{fig:debiased_aei_pha}), even at the lowest inclinations (the histograms have been normalized by their sums).  However, the PHAs with higher absolute magnitudes (which are likely to be smaller) show an overabundance of low inclination objects that is similar to that observed in the NEOWISE sample. Whether or not this result is due entirely to the survey biases of the ground-based surveys that have discovered the preponderance of PHAs remains to be determined.}
\end{figure}

We can compare this result from NEOWISE alone to the inclination distribution of all currently known PHAs. Such analysis is complicated by the fact that most PHAs do not have well-known diameters, only $H$, and by the fact that unlike the NEOWISE survey, we have not accounted for the biases of the surveys that have discovered most of the objects.  With these caveats, we can compare the inclination distributions of the known PHAs with $H<18$ mag to those with $H>18$ mag (Figure \ref{fig:known_pha_incl_hist}).  We can make the assumption that the sample of known PHAs with $H<18$ mag is roughly complete, and we see that relative numbers of objects in the first two inclination bins roughly resembles that of the S3M model (Figure \ref{fig:debiased_aei_pha}).  Yet the inclination distribution of the known PHAs with $H>18$ mag shows an overabundance of low inclination objects similar to that observed in the NEOWISE sample.  However, caution must be used when interpreting this result: unlike the all-sky WISE survey, which we have shown to be largely unbiased with respect to inclination, the ground-based surveys that have discovered most of the known PHAs spend more of their time surveying near low inclinations, so the observed overabundance of low inclination objects could be the result of survey biases.  Nonetheless, it is possible that the overabundance of small, low inclination PHAs with respect to the S3M model is echoed in the entire sample of known PHAs.  A logical next step would be to determine the survey biases for the ground-based surveys and see whether or not the overabundance of small, low inclination PHAs relative to the S3M model remains after they have been removed. 

\begin{figure}
\figurenum{6}
\includegraphics[width=3in]{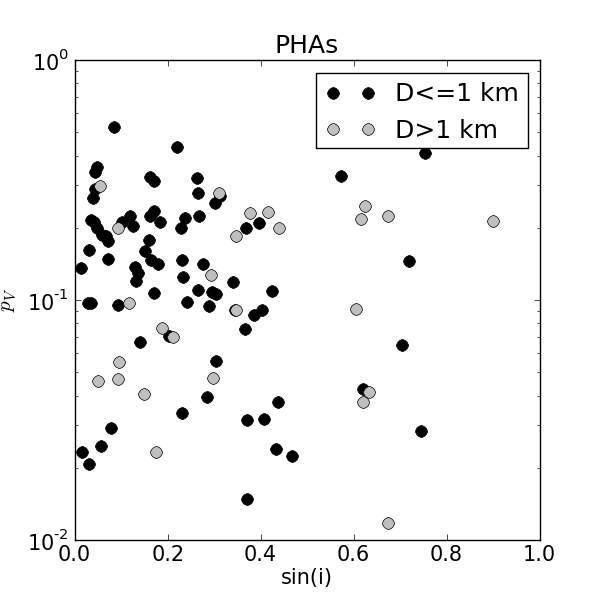}
\caption{\label{fig:pv_incl} Low-inclination PHAs detected by NEOWISE during the fully cryogenic portion of the mission may be somewhat smaller and have higher albedos than those with higher inclinations, although the sample size is small.  The figure shows visible albedo vs. inclination for PHAs grouped by diameter (larger and smaller than 1 km). The NEOWISE survey has been shown to detect asteroids with relatively little bias in either visible albedo and inclination, so the plot is likely representative of the larger population.}
\end{figure}

Unlike most of the known PHAs, for which only $H$ is known, we have determined the sizes and albedos of the NEOWISE PHA sample, as well as the survey biases.  Using this information, we can consider whether or not the overabundance of low inclination PHAs correlates with their sizes or albedos.  While we have shown in M11b that NEOs as a whole have no significant change in their albedos with diameter, Figure \ref{fig:pv_incl} shows that the small, low inclination PHAs may be somewhat brighter than their larger counterparts; a two sample Kolmogorov-Smirnov test comparing the albedo distributions of PHAs larger and smaller than 1 km with $sin(i)<0.2$ yields a test statistic $D=0.60$ and probability $P=0.0056$.  

The observed difference in the inclination distributions of large and small PHAs might be unexpected from dynamical considerations alone.  Gravitational perturbations act independently of an object's size, and the Yarkovsky force (which does depend on diameter) does not tend to drive objects to lower inclinations \citep{Vokrouhlicky.1998a}.  NEO orbital elements evolve on timescales of $\sim$1-100 Myr \citep{Morbidelli.1998a}, and in some cases significantly faster \citep[e.g.][]{Michel.1996a}, so one would expect that the large and small PHAs would have nearly identical inclination distributions, assuming that they originated from the same source regions in similar proportions.  However, is is possible that the PHAs may not all originate from the same sources.  For example, the Flora family spans the $\nu_{6}$ resonance thought to be a key source region for the NEOs \citep{Bottke.2002a}. It has a low mean inclination \citep[$\sim 4^{\circ}$;][]{Hirayama.1918a}, and  \citet{Masiero.2011a} showed that most Flora family members are characterized by high albedos.  \citet{Bottke.2005a} show that breakups of 10 km asteroids occur approximately every $10^{5}$ years in the Main Belt.  If Flora is $\sim$1 Gyr old \citep{Nesvorny.2002a}, and there are $\sim$55 Flora members larger than 10 km \citep{Masiero.2011a}, then approximately half of these objects should have disrupted over the age of the family, or one breakup every $\sim$30-40 Myr.  If the timescale for resupplying the NEOs is $\sim$10 Myr, then there is a $\sim$1/3 chance that such a breakup is responsible for the present-day makeup of the NEOs.  Yet a 10 km breakup is not needed in order to produce the numbers of small, sub-km fragments that we observe in Figure \ref{fig:pv_incl}; a single breakup would be expected to produce several hundred pieces $\sim$1/10th the size of the parent \citep[c.f.][]{Durda.2007a}. A $\sim$1-2 km parent could produce several hundred 100 m fragments.  The Flora family contains several thousand members with $D>1$ km \citep{Masiero.2011a}. \citet{Bottke.2005a} predicts that 1 km asteroids should disrupt every 300 Myr, so with $\sim$4000 members, a 1 km Flora family member should break up approximately every 75,000 years.  Thus, a surplus of low inclination PHAs that we observe today could be a ``snapshot" of the aftermath of a stochastic event in the Main Belt. In another $\sim10^{5}$ years, the inclination distribution of the smaller objects would tend to be dynamically reshuffled, or perhaps resupplied by another breakup.  Although this speculation is by no means conclusive proof that the low inclination PHAs originate within the Flora family, it illustrates the feasibility of such an occurrence.   

Our results would be strengthened by acquiring a much larger sample of PHAs detected with a uniform (or at least well-known) bias in inclination.  Future work will include lowering signal-to-noise thresholds and the number of required detections in order to find more asteroids within the NEOWISE dataset.  Nevertheless, the result that there may be more PHAs with low inclination than the S3M model predicts may indicate that there are potentially more objects with low $\Delta v$ with respect to the Earth, and that there may be more objects with a potential for Earth impacts in the next 100 years.  The larger number of low-inclination PHAs could also give rise to more of the small (meter-sized) objects predicted to be temporarily captured by the Earth \citep{Granvik.2011a}. It should be noted that the S3M model does not yet include any ``horseshoe" objects or Earth Trojans such as 2010 TK$_{7}$ \citep{Connors.2011a}, which was discovered by NEOWISE after the depletion of all cryogen.  Such classes of objects could also be more likely to have low $\Delta v$ and a greater chance of becoming impactors than the average NEO.  

In order to revise the actual impact hazard to the Earth, we will need to revise the model of PHAs and numerically integrate to see how many objects are likely to become impactors.  \citet{Ito.2010a} argued that the asymmetry in younger, rayed craters observed on the Moon reported by \citet{Morota.2003a} is the result of a hitherto undetected population of NEOs in highly Earth-like orbits whose average impact velocities on the Moon are much lower than the average impact velocity of the NEOs produced by the \citet{Bottke.2002a} model.  They carried out numerical integrations to simulate the orbital evolution of test particles with orbits generated from the \citet{Bottke.2002a} model and concluded that the model underpredicts the number of NEOs with low inclinations and semimajor axes near 1 AU, resulting in low $\Delta v$ with respect to the Moon.   Our result, that there is a factor of $\sim$2.3 more PHAs with $\sin(i)<0.1$ than predicted by the \citet{Bottke.2002a} model (which will tend to have lower impact velocities), may represent the missing population predicted by \citet{Ito.2010a}.  As noted above, our result does not yet include models of the populations of Earth Trojans and horseshoe objects, which would also tend to have lower impact velocities relative to Earth and the Moon.  Future work will study the impact hazard to the Earth and the Moon through numerical studies of the model population's orbital evolution.

\subsection{Atens, Apollos, Amors}
As discussed above, Figures \ref{fig:orbital_elements_allNEAs} and \ref{fig:neo_types1} illustrate that there is a correlation between \pv, $D$, and the orbital elements of the NEAs.  In order to more accurately account for these correlations in our simulations, as a preliminary step, we split the NEOs into three groups: the Atens, Apollos, and Amors.   Following a similar method to the technique described above for the PHAs, we computed the survey biases for each of the three groups and use these biases to study the physical and orbital properties of each population.  

\begin{figure}
\figurenum{7}
\includegraphics[width=6in]{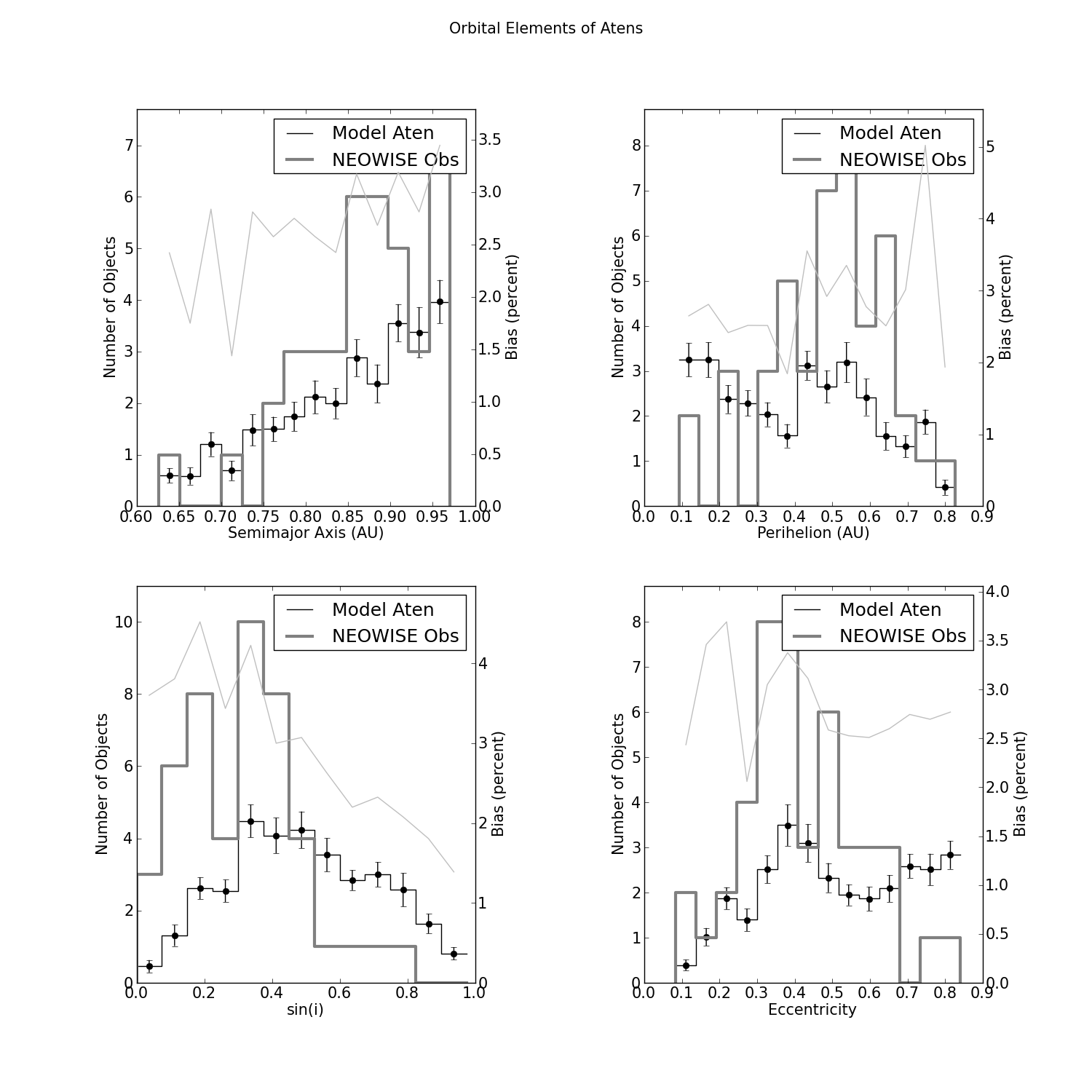}
\caption{\label{fig:debiased_aei_aten} Each panel shows the orbital elements of the Atens observed by NEOWISE during the fully cryogenic portion of the mission (medium gray lines) compared to the number predicted by computing the survey biases (light gray lines) and applying them to the S3M model (black lines).  Error bars on the model were determined through Monte Carlo simulations.}
\end{figure}

Computing the NEOWISE survey bias for Atens with $D>100$ m ($P_{N}$ in Equation 2 above) and applying it to the Atens detected by NEOWISE during the fully cryogenic portion of the survey yielded an estimated total of 1600$\pm$760 larger than 100 m and 42$\pm$31 larger than 1 km. The best-fitting slope of the power law representing the cumulative size frequency distribution of the Atens between 100 m and 2.2 km is $\alpha=1.63\pm 0.30$.  Figure \ref{fig:debiased_aei_aten} shows a comparison between the orbital elements of the model objects that were found by the simulated survey compared with the Atens actually detected by NEOWISE.  The S3M model appears to predict a factor of 6.5 times fewer low inclination objects ($\sin(i)<0.1$) and an overabundance of high inclination objects relative to the observations, although the significance is low (2.5$-\sigma$ for low inclination) due to the size of the sample.  NEOWISE discovered one Aten larger than 1 km, which results in 3$\pm3$ Atens larger than 1 km remaining to be found; therefore, 92\% of the population of large Atens have been discovered.  

\begin{figure}
\figurenum{8}
\includegraphics[width=6in]{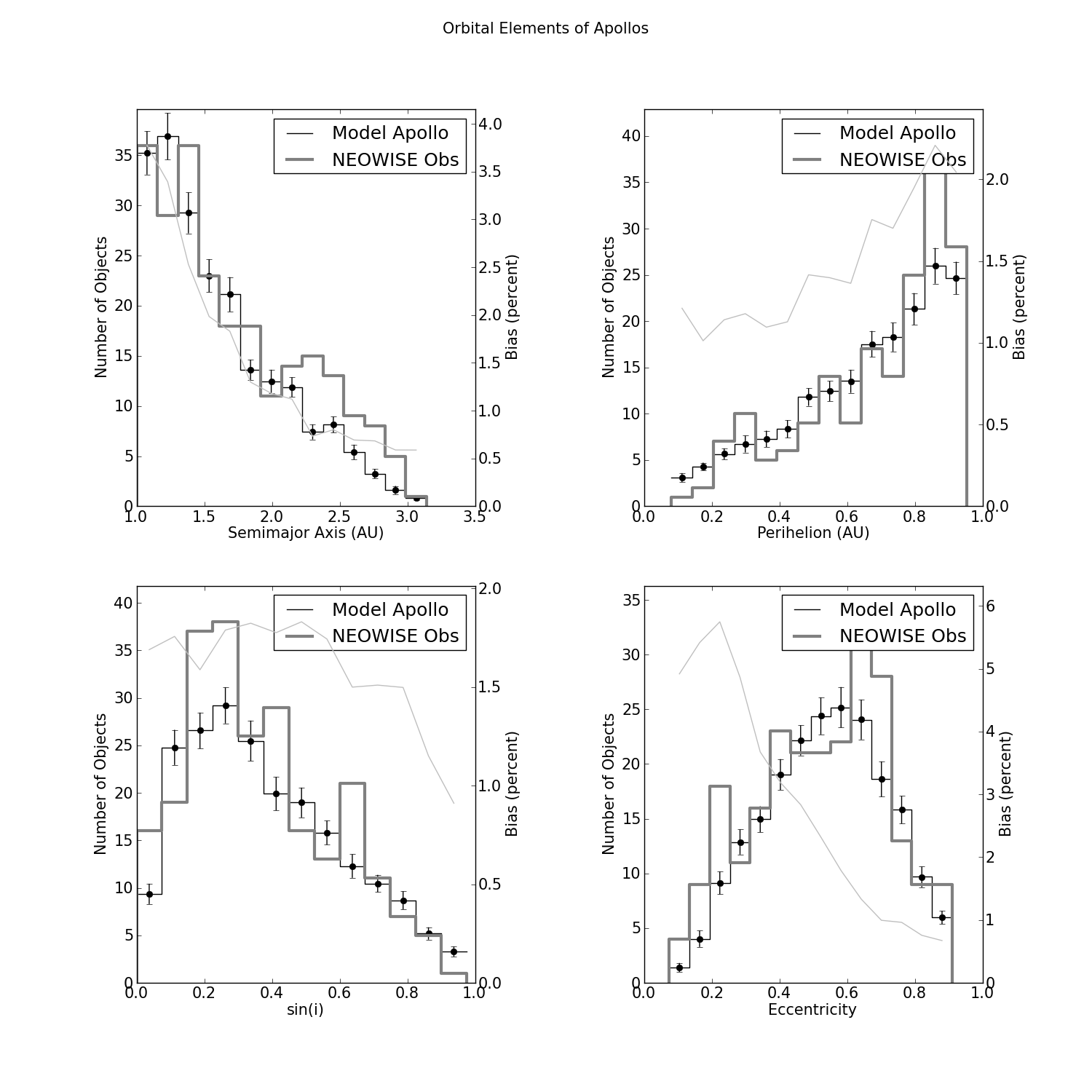}
\caption{\label{fig:debiased_aei_apollo} Each panel shows the orbital elements of the Apollos observed by NEOWISE during the fully cryogenic portion of the mission (medium gray lines) compared to the number predicted by computing the survey biases (light gray lines) and applying them to the S3M model (black lines).  Error bars on the model were determined through Monte Carlo simulations.}
\end{figure}

Using the same methods described above, we find that there are 462$\pm$110 Apollos larger than 1 km and 11200$\pm$2900 larger than 100 m.  The Apollos' cumulative size frequency distribution was determined through a least-squares fit to be best represented by a broken power law with a slope $\alpha=1.44\pm 0.12$ for 100 m $< D <$1.6 km and $\alpha=3.0\pm 1.5$ for $D>1.6$ km.  Figure \ref{fig:debiased_aei_apollo} shows the orbital element distributions of the model predictions compared to the observations for the Apollos, and they are generally in good agreement.  Although the model predicts 70\% fewer Apollos with $\sin(i)<0.1$, the statistical significance is $\sim$2.5$-\sigma$.  There were five Apollos larger than 1 km discovered by NEOWISE, and computing and removing the survey biases using the approximation that $P_{N_{disc}}\rightarrow P_{N}$ results in an estimate that there are 30$\pm$19 Apollos larger than 1 km remaining to be discovered.  Therefore, approximately 93\% of the Apollos larger than 1 km are likely to have been found to date.  

\begin{figure}
\figurenum{9}
\includegraphics[width=6in]{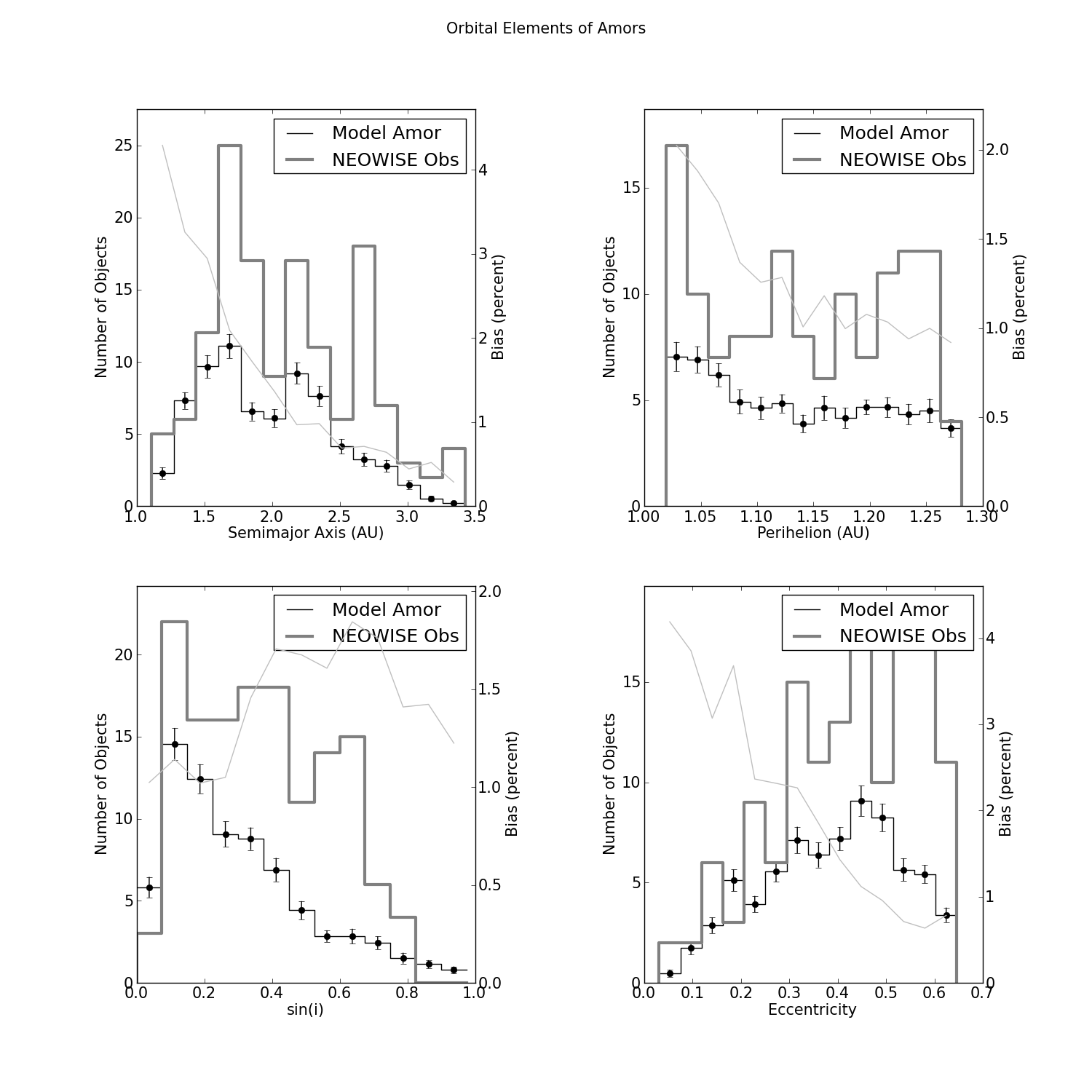}
\caption{\label{fig:debiased_aei_amor} Each panel shows the orbital elements of the Amors observed by NEOWISE during the fully cryogenic portion of the mission (medium gray lines) compared to the number predicted by computing the survey biases (light gray lines) and applying them to the S3M model (black lines).  Error bars on the model were determined through Monte Carlo simulations.}
\end{figure}

By creating a synthetic population of Amors and using Equation 2 to estimate and remove the survey biases from the observed objects, we compute that there are 320$\pm$90 Amors larger than 1 km and 7700$\pm$3200 larger than 100 m.  The functional form of the cumulative size frequency distribution for the Amors is best fit by a broken power law with $\alpha=1.40\pm0.18$ for 100 m $< D <$1.9 km and $\alpha=5.0\pm2.0$ for $D>1.7$ km.  NEOWISE discovered 10 Amors larger than 1 km, and as we are reasonably certain that nearly all NEOs in this size range have been found, we can use the assumption that the survey biases of the discovered objects are the same as the biases for all Amors larger than 1 km detected by NEOWISE.  Applying these biases to the NEOWISE discoveries larger than 1 km yields an estimated 51$\pm$30 that remain to be found.  This result suggests that the completeness of the Amors is $\sim83\%$, somewhat less than that of the NEOs overall.  The comparison of the NEOWISE-observed Amors' orbital elements to the S3M model predictions is shown in Figure \ref{fig:debiased_aei_amor}.   In the case of the Amors, the model predicts an overabundance of low-inclination Amors compared to the observed population of a factor of nearly two, with a statistical significance of $\sim2-\sigma$ assuming that the number of objects detected by NEOWISE follows Poissonian statistics.  However, the model underestimates the number of Amors with inclinations in the range $\sin(i)=0.3-0.75$, producing about a factor of two fewer objects with a $\sim5-\sigma$ statistical significance.  As \citet{Bottke.2002a} suggests that the Amors are preferentially produced from cometary populations and are likely to have been heavily influenced by Jupiter, this result suggests that the model may not adequately capture the degree to which the Amors' inclinations has been pumped up by such encounters for these objects.  

\begin{figure}
\figurenum{10}
\includegraphics[width=6in]{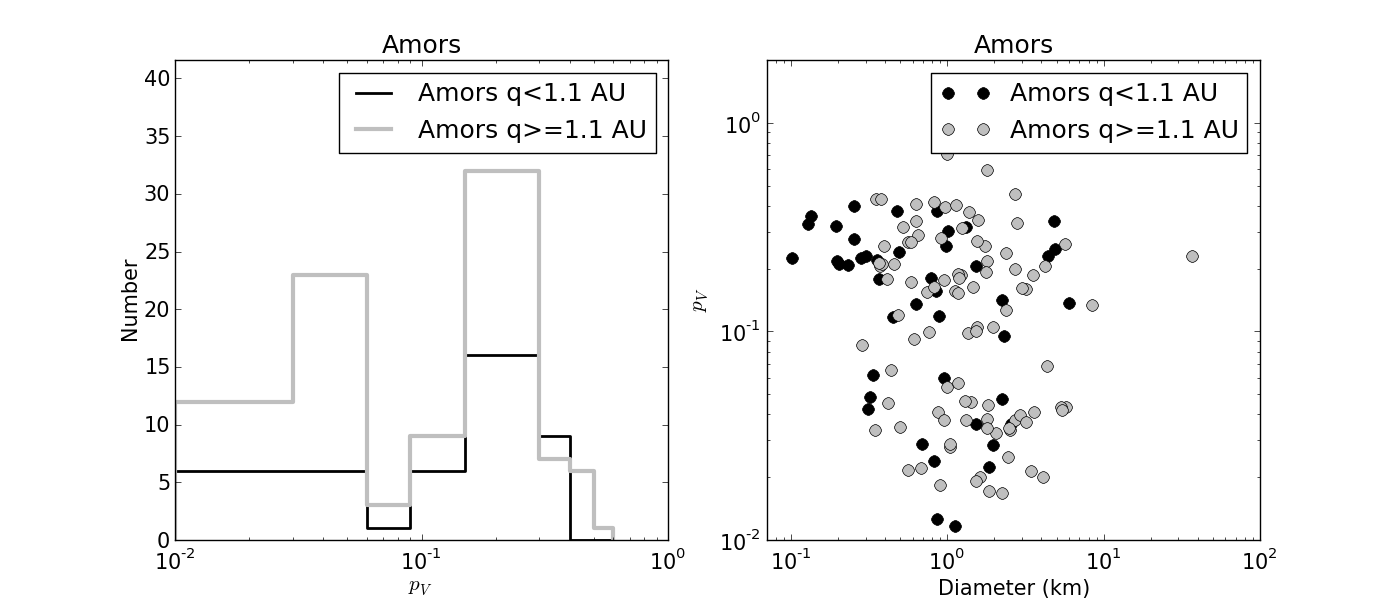}
\caption{\label{fig:amor_q} Separating the Amors into those with q$<$1.1 AU and those with a$\geq$1.1 AU reveals that the two groups may have somewhat different physical properties; Amors with q$<$1.1 AU may be brighter, possibly reflecting different origins. }
\end{figure}

We now consider separating the Amors into two groups, those with perihelia above or below 1.1 AU, since as discussed above, $q=1.1$ AU may represent a boundary between Amors with different origins.  Figure \ref{fig:amor_q} shows the relationship between albedo and diameter for these two types of Amors.  While in M11b we showed that there is no correlation between visible albedo and diameter for the NEOs as a whole, Figure \ref{fig:amor_q} shows that Amors with $q<1.1$ AU may be somewhat brighter at smaller sizes than those with $q>1.1$ AU, although the sample size is small.  Even though the NEOWISE sample is essentially unbiased with respect to visible albedo, Amors with small diameters and $q>1.1$ AU were less likely to have been detected by WISE, complicating our ability to study the albedo distribution of small vs. large Amors with $q>1.1$ AU.  This result may support the notion that the high perihelia Amors are more likely to be dark, dead or dormant comets that have been scattered inward by Jupiter, and the low perihelia Amors are more likely to originate from a different, higher albedo source region.  

By separating the NEAs into three subpopulations and determining the survey biases for each separately, we have obtained a refined estimate of the total numbers and sizes compared to M11b, which treated the NEAs as a single group.  Combining the totals of Atens, Apollos, and Amors derived using the survey debiasing techniques described above produces 824$\pm$145 NEAs larger than 1 km and 20,500$\pm$4200 larger than 100 m, both within 1-$\sigma$ of our original estimates in M11b.   As noted above, these numbers can be regarded as a lower limit of sorts; as more of the lost NEOWISE-discovered objects are recovered, the sample size will increase somewhat.  Our results indicate that the fractions of Atens, Apollos, and Amors in the total NEO population are somewhat different than the predictions of \citet{Bottke.2002a}, being 8$\pm$4\%, 55$\pm$18\%, and 37$\pm$16\% respectively.  

Figure \ref{fig:neo_types1} compares the albedo distributions between the different types of NEOs.  A two-sample Kolmogorov-Smirnov  test comparing the albedo distributions of the Atens and the Amors with perihelion $>$1.1 AU yields a test statistic $D=0.32$ and a probability $P=0.002$.  This calculation suggests that it is 99.7\% likely that the two albedo distributions are not drawn from the same population, a 3$-\sigma$ result.  The Atens and the NEOs with $T_{J}\leq3$ are the most dissimilar; the figure shows that the Atens and Apollos have somewhat similar albedo distributions ($D=0.29$; $P=0.017$), and the Apollos and Amors have more similar albedo distributions ($D=0.09$; $P=0.399$).  The PHA albedo distribution has the highest probability of being drawn from an Apollo-like population ($D=0.05$; $P=0.940$ vs. $D=0.26$ and $P=0.019$ for the Atens and $D=0.11$ and $P=0.388$ for the Amors).  Although this result suggests that the PHAs are primarily drawn from the Apollos, the actual hazard posed by the PHAs depends on the number of chances  each object has of impact over time, so the Atens among them may pose more risk overall.

\begin{figure}
\figurenum{11}
\plottwo{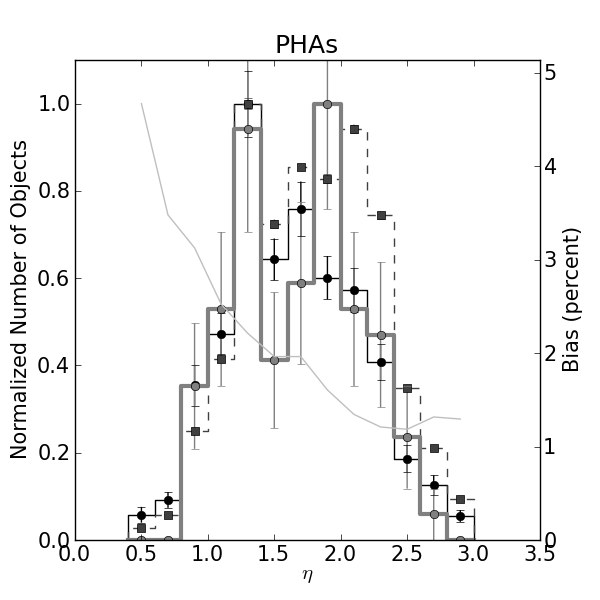}{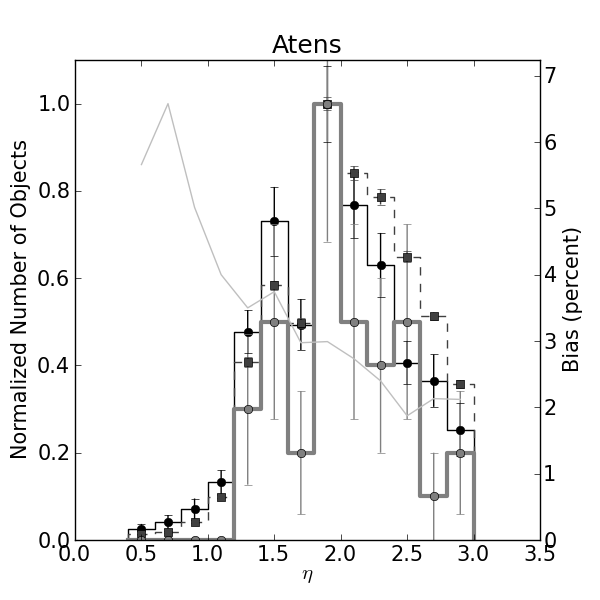}
\plottwo{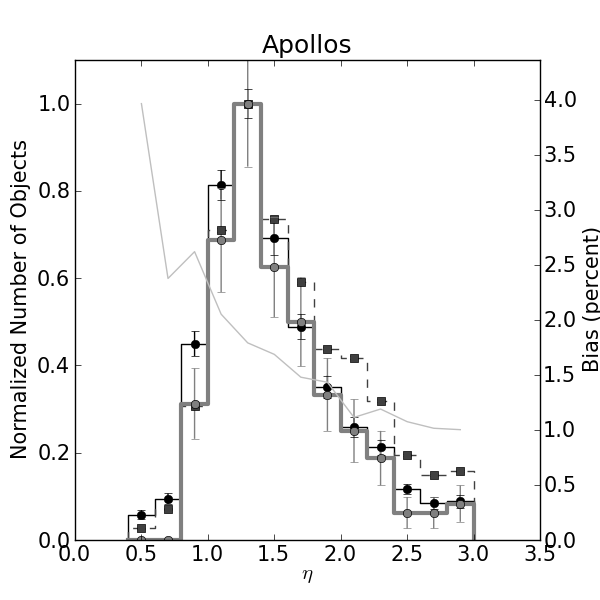}{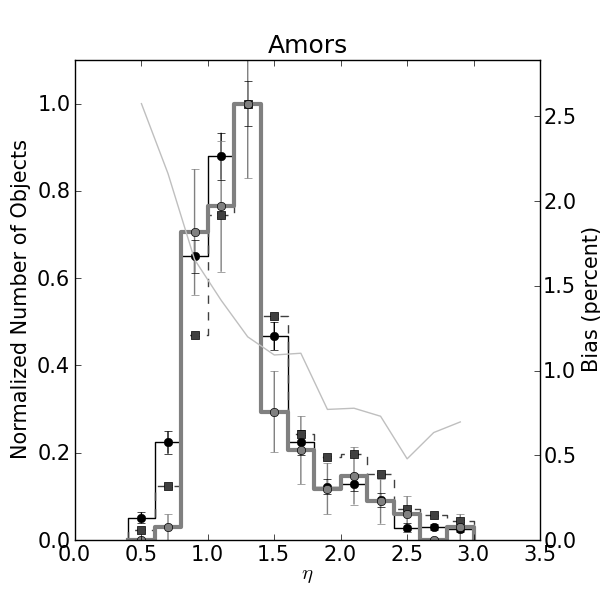}
\caption{\label{fig:beaming} Each panel shows the beaming distribution of the PHAs (top left), Atens (top right), Apollos (bottom left), Amors (bottom right) observed by NEOWISE during the fully cryogenic portion of the mission (medium gray lines) compared to the number predicted by computing the survey biases (light gray lines) and applying them to the S3M model (black lines).  The dashed line represents the synthetic population.  Error bars on the model were determined through Monte Carlo simulations.  The beaming distribution of the four populations is different, but as the Atens and PHAs were observed at significantly higher phase angles than the Apollos and Amors, this result could be entirely due to the correlation between $\eta$ and phase angle/heliocentric distance for NEOWISE-observed NEOs rather than real thermophysical differences.}
\end{figure}

Figure \ref{fig:beaming} shows a comparison of the beaming parameter ($\eta$) distributions of the PHAs, Atens, Apollos, and Amors.  The Amors tend to have lower $\eta$ values than the Apollos, which in turn have lower $\eta$ values than the Atens and the PHAs.  The beaming parameter is thought to be related to thermophysical properties such as thermal inertia and surface roughness \citep[c.f.][]{Harris.2009a,Delbo.2007a}; however, as shown in M11b, $\eta$ is strongly correlated with phase angle and heliocentric distance for NEOs observed by WISE since WISE was constrained to observe near 90$^{\circ}$ solar elongation.  NEATM assumes that an asteroid's night contributes negligible thermal flux, yet WISE observed most NEOs at higher phase angles and therefore usually observed some fraction of the objects' night sides.  If the night side thermal flux is not actually negligible (because temperature does not fall to near-zero on the night side), then NEATM will fail to predict the correct flux to varying degrees depending on how much of the night side is observed.  Although one might conclude that lower $\eta$ values imply lower thermal inertia for the Amors, the relationship between $\eta$ and phase angle complicates efforts to link it directly to such physical properties.  The Amors tended to be observed by WISE at lower phase angles than either the Apollos or Atens (mean phase angles were 46$^{\circ}$, 58$^{\circ}$, and 66$^{\circ}$, respectively), so the generally lower $\eta$ values for the Amors could be due entirely to the correlation between $\eta$ and phase angle/heliocentric distance. A thermophysical model should be used instead of NEATM to compute thermal inertia directly if the rotational state can be determined. 

\subsection{Cometary Bodies Among the NEOs}
The question of how many of the NEOs are actually of cometary origin (both active comets and dead or dormant comets) is complicated, and the estimated fraction is thought to be somewhat less than 10\% \citep{DeMeo.2008a,Fernandez.2005a}.  In \citet{DeMeo.2008a}, dormant comet candidates among the NEOs were defined as objects with Jupiter Tisserand invariant $T_{J}<3$ that either had albedos less than 9\% or spectral types C, P, or D. \citet{Fernandez.2005a} estimated that 4\% of the NEOs are actually dormant comets.  In the fully cryogenic mission, NEOWISE detected 46 NEOs with $T_{J}\leq 3$.  It should be noted that this list does not include known near-Earth comets that were observed or discovered by NEOWISE.  However, not all objects with $T_{J}\leq3$ are actually cometary in origin; many could have migrated from the inner, middle or outer portions of the main asteroid belt through mean motion resonances, as noted in \citet{Bottke.2002a}. 

Having $T_{J}\leq3$ should not be taken to universally mean that an object is actually a dormant comet.  Most of the objects with $T_{J}\leq 3$ have very high inclinations and eccentricities, as expected for NEAs.  Some of these NEOs may have evolved from Main Belt orbits near the $\nu_{6}$, 3:1 or 4:1 mean motion resonances with Jupiter into Earth-crossing orbits, for example.  This possibility is strengthened by examining their albedo distribution (Figure \ref{fig:neo_types1}), which reveals that 50\% of the NEAs with $T_{J}\leq3$ have \pv$\geq$10\%, while 62\% of the NEAs with $T_{J}>3$ have \pv$\geq10\%$.  Some of these low Tisserand, high albedo NEOs may not be cometary in origin at all, while others could be active comets with activity that was unresolved in the WISE images or in the ground-based images used to determine absolute magnitude. For example, by its orbital elements, 2009 UV18 appears to be a comet (it has semimajor axis $a=3.18$ AU, eccentricity = 0.63, and inclination of 8$^{\circ}$), yet its visible albedo is 71\%. Although it is likely that bad $H$ or $G$ values for this object are at least partially responsible for its high albedo, deep optical imaging of this $T_{J}=2.8$ NEO and others like it to search for activity would help to understand its nature and origin, along with numerical studies of its orbital evolution. 

Of the 46 NEAs observed by NEOWISE with $T_{J}\leq3$, there are 20 with \pv$<$0.075 (the criteria used by \citet{Fernandez.2005a} to classify an object as a dormant comet candidate). Thus, about 43\% of these NEAs detected by NEOWISE satisfy these criteria, representing 5\% of the current sample of 429 NEAs.  As previously stated, the list of 46 NEAs with $T_{J}\leq3$ does not include known near-Earth comets that NEOWISE observed during the fully cryogenic mission. Of the 93 active comets detected by the automated WMOPS portion of NEOWISE pipeline to date, seven are near-Earth comets, or 7.5\%. Analysis of the active near-Earth comets is ongoing, and their cumulative size distribution cannot be properly evaluated and debiased until accurate nuclear diameters can be determined. Determining the true fraction of active and dormant comets within the NEOs requires more detailed numerical studies of their orbital evolution, estimates of their nuclear sizes, and deep imaging to understand their activity; this will be addressed in a future work.  

\section{Conclusions}
We have shown that the various subpopulations within the near-Earth objects encompass a diverse array of albedos and cumulative size distributions.  We have provided an estimate of the number of potentially hazardous asteroids; the characteristics of this population in particular should be evaluated when computing likely future impact risks.  We have shown that more than 90\% of the PHAs larger than 1 km have already been discovered.  Approximately 30\% of the $\sim4700\pm$1450 PHAs larger than 100 m have been discovered.  While we found in M11b that the overall orbital element distribution model of \citet{Bottke.2002a} matches the NEOs as a whole reasonably well, there are exceptions within certain subpopulations, including the PHAs.  We have shown that the slope of the cumulative size distribution of the PHAs is somewhat steeper than that of the NEAs, although the difference in the slopes of the two groups is within a single standard deviation.  The Atens have brighter albedos than the Amors, and their beaming distributions are different, possibly highlighting their diverse origins.  The PHAs are somewhat brighter than the NEOs.  The Atens and Apollos most likely originate within the brighter regions or collisional families of the Main Belt, and the Amors are more likely to have evolved from darker source regions in the Main Belt or be cometary bodies scattered inward by Jupiter.  We find that there are approximately twice as many high-inclination Amors as the S3M model predicts.  Since the Amors are rarely PHAs, they are not the dominant population of interest when considering the hazard from near-Earth objects.  The Amors themselves can be divided into those with perihelia above and below 1.1 AU; the group with q$<$1.1 AU appears to be somewhat brighter than the group with larger perihelia, suggesting that the two groups may have different origins.   

By itself, a measurement of a population's size frequency distribution slope only yields information as to whether or not that system is in collisional equilibrium if the system is self-similar \citep[e.g.][]{Dohnanyi.1969}.  With their wide range of albedos, taxonomic types, and compositions, the NEAs cannot reasonably be assumed to be self-similar.  The significance of the varying size distribution slopes of the Atens, Apollos, and Amors that we have found can only be assessed in the context of a larger model that accounts for the system's full collisional evolution and physical properties.  \citet{Bottke.2002a} found a slope of $\alpha=1.75\pm0.1$ for the NEO cumulative size distribution between 200 m and 4 km in diameter, and they assumed that the slope was the same for all NEO types.  \citet{Obrien.2005a} used the observed NEO and Main Belt asteroid cumulative $H$ distributions to study the collisional evolution of the asteroids and place constraints on their strength parameters.  The differing slopes of the cumulative size distributions from 100 m to 1 km of the Atens, Apollos and Amors may be indicative of different compositions, densities, or collisional histories, but specific conclusions on this front must await more detailed modeling and comparison to the Main Belt and comets.

Finally, we find that the current best model of the PHAs underestimates the number of low-inclination objects by a factor of $\sim$2.3$\pm0.7$, although we caution that the sample of low-inclination PHAs detected by NEOWISE is small.  This result suggests that the actual impact hazard from PHAs may be somewhat more than expected from consideration of their numbers alone because low inclination PHAs will tend to have more chances to impact the Earth.  The change in impact hazard (i.e. the number of impacts that will occur with a certain energy over time) will be the subject of future work based on an analysis of a revised model's orbital element evolution.  The increased number of PHAs compared to the \citet{Bottke.2002a} model furthermore suggests that while the overall number of NEAs may be fewer than predicted (M11b), there may be more of the objects that are the most energetically accessible to future missions.  The low inclination PHAs smaller than 1 km in the NEOWISE sample may be somewhat brighter than the PHAs larger than 1 km.  While these results are difficult to explain dynamically, it is possible that the low inclination PHAs may represent a population with different origins.  As an example, we show that the excess of low inclination PHAs could have been supplied by the breakup of a 1-2 km object in a high albedo Main Belt family near a resonance.  

The statistical significance of our results would be improved by the addition of more objects with well-determined diameters and albedos that have been detected with well-understood survey biases.  Future work will include reprocessing the WISE data to extract more NEOs in a systematic fashion using an improved version of the WMOPS algorithm and the optimized set of image calibration products that have recently become available, as well as data from the post-cryogenic portion of the survey.  Our results are still subject to additional uncertainties due to the candidate NEOs that appeared on the Minor Planet Center's NEO Confirmation Page but received neither ground-based follow-up nor designations, as well as those with arcs too short to have been placed on the Confirmation Page.  Some of the WISE discoveries may not have been correctly classified as NEOs due to their short observational arcs.  Any additional objects found by making linkages to other survey data will increase the size of the estimated populations.  It would also be desirable to incorporate recent data from the ground-based visible surveys, which requires a detailed accounting of their survey biases \citep[c.f.][]{Bottke.2002a,Spahr.1998a,Jedicke.1998a}.  At that time, the sub-populations within the NEOs will be revisited. 

\section{Acknowledgments}

\acknowledgments{This publication makes use of data products from the \emph{Wide-field Infrared Survey Explorer}, which is a joint project of the University of California, Los Angeles, and the Jet Propulsion Laboratory/California Institute of Technology, funded by the National Aeronautics and Space Administration.  This publication also makes use of data products from NEOWISE, which is a project of the Jet Propulsion Laboratory/California Institute of Technology, funded by the Planetary Science Division of the National Aeronautics and Space Administration.   We thank our referee, Dr. Alan Harris of the Space Sciences Institute, for his helpful comments, which lead to a number of new insights, in particular the suggestion to split the Amors by perihelion.  We gratefully acknowledge the extraordinary services specific to NEOWISE contributed by the International Astronomical Union's Minor Planet Center, operated by the Harvard-Smithsonian Center for Astrophysics, and the Central Bureau for Astronomical Telegrams, operated by Harvard University.  We also thank the worldwide community of dedicated amateur and professional astronomers devoted to minor planet follow-up observations. This research has made use of the NASA/IPAC Infrared Science Archive, which is operated by the Jet Propulsion Laboratory, California Institute of Technology, under contract with the National Aeronautics and Space Administration.}

  \clearpage

 \clearpage
 
\end{document}